\documentclass[fleqn,12pt]{article}
\usepackage[numbers,sort&compress]{natbib}
\usepackage[footnotesep=0.4in]{geometry}
\usepackage{graphicx}
\usepackage{caption2}
\usepackage{amsmath}
\usepackage{bbding}
\usepackage{graphics}
\usepackage{pifont}
\usepackage{amsfonts}
\usepackage{color}
\usepackage{caption2}
\usepackage{float}
\usepackage{mathrsfs}
\usepackage{amssymb}
\usepackage{epstopdf}
\usepackage{appendix}
\graphicspath{{./image/}} \makeatletter
\makeatletter

\newcommand\figcaption{\def\@captype{figure}\caption}
\newcommand\tabcaption{\def\@captype{table}\caption}
\begin{document}

\date{}
\title{Particle migration in porous media: from the mesoscopic perspective}
\author{Sheng Zhang$^1$, Tong Zhang$^1$, Jing-Jing Su$^1$\thanks{Corresponding
author, with e-mail address as jingjingsu@csu.edu.cn}, Dai-Chao Sheng$^{1,2}$\\ \\
{\em 1. School of Civil Engineering, Central South University, Hunan 410075, China}\\
{\em 2. School of Civil and Environmental Engineering, University of}\\
{\em Technology Sydney, Sydney 2007, Australia}
}

%Nonlinear wave study of certain intractable shallow water wave equations via the physics-informed neural networks
%
%Explorations on nonlinear waves for certain intractable shallow water wave equations via the physics-informed neural networks

\maketitle
%\vspace{-7mm}
\begin{abstract}
Convection-diffusion equation is used to describe particle migration process in many fields, while it is proposed based on the empirical Fick's law. In this paper, with the help of the percolation model, we theoretically investigate the particle migration law in porous media from the mesoscopic perspective, and base on the probabilistic migration characteristic of particles to strictly reformulate the convection-diffusion equation. Meanwhile the quantitative relations between the convection, diffusion coefficients and the mesoscopic parameters of particle-motion and percolation-configuration are revealed. Furthermore, via the Monte-Carlo numerical simulation, we verify the proposed mesoscopic particle migration theory and modify the expressions of convection, diffusion coefficients for global applicability. In addition, applicable qualification of the proposed mesoscopic theory is given, and relation between the blocking effect parameter and connecting probability of the percolation model is obtained.

\end{abstract}

\vspace{3mm}

\vspace{20mm}

%\newpage
%\textbf{Boussinesq and Camassa-Holm equations are respectively used to describe the bidirectional and unidirectional motions of small-amplitude waves on the shallow water surfaces, which are both numerically intractable problems due to the instability of the equation or the irregularity of the solutions. In this paper, for the two equations, via the physics-informed neural networks (PINNs), we achieve the meaningful problems for predicting the soliton (peakon) interaction or rogue wave behaviors accurately with few initial and boundary data, which demonstrates that the PINNs may
%provide an effective way to deal with certain unstable physical systems or extreme solutions of regular systems. Furthermore, we propose multiple PINNs to reflect the wave-depth parameters stably based on the known wave behaviors. Compared with the conventional numerical methods, the PINNs are revealed to present higher-precision wave dynamics with fewer initial and boundary data, which can be extended into more physical systems for discovering new phenomena.
%}
\newpage
\noindent {\Large{\bf 1. Introduction}}\vspace{3mm}

%presents a straightforward description, exhibit, present
%plays an important role, reflect, capture, express, achieve
%introduced, used, model considered, simulate,straightforward,determine, developed from
% As so many different substances show similar behavior, there may be a general explanation for these phenomena and a common perspective to understand and deal with related problems~\cite{}.  in the subsurface

Particle migration in porous media is a universal physical phenomenon appearing in nature, such as subsurface organic carbon's permeation, groundwater pollutant migration and aerosol particles' diffusion in atmosphere~\cite{Reichstein,Kersting,Hinds}. In the aspect of particle migration rule, convection-diffusion equation is widely used to describe a variety of particle migration processes~\cite{Chandrasekhar,Bird}. As is well known, convection-diffusion equation has been derived based on the macroscopic Fick's law, i.e., ``the molar flux due to diffusion is proportional to the concentration gradient"~\cite{Davis}. Indeed, Fick's law well fits the experimental data, while there is no rigorous theoretical proof for it up to now, as is the convection-diffusion equation~\cite{Davis}.
%basis,simulating this law for generating percolation models

For investigating porous media theoretically, percolation models have been introduced based on the percolation theory~\cite{Schaap}. Percolation model is a regular space lattice structures composed of a series of vertices or edges, and every vertex or edge is occupied to be open with a certain probability, via the two adjacent occupied vertices or the occupied edge, such tiny matter as particles and microorganism can pass through~\cite{Broadbent,Christensen}. With different occupied probabilities, the percolation model may present diametrically opposed performances of connectivity~\cite{Broadbent,Christensen}. At present, there are
many percolation models, including isotropic percolation, directed percolation, invasion
percolation, first-passage percolation, explosive percolation, etc.~\cite{Broadbent,Christensen}. Recently, percolation theory has been used in such fields as physics, biology, and geophysics for interpreting the corresponding phenomena~\cite{Schaap,Broadbent,Christensen}.

As we know, convection and diffusion are two common phenomena in porous media, percolation theory is also closely related to porous media, while there are little rigorous theoretical work discussing the relation between those two fields.
In this paper, with the help of the percolation model, we will investigate the particle migration law in porous media, and reformulate the convection-diffusion equation without using the Fick's law from the mesoscopic perspective. This paper will be organized as follows. In Section 2, by reducing a symmetrical three-dimensional bond percolation model into the mesoscopic one-dimensional particle migration model, we will derive the convection-diffusion equation based on the probabilistic migration characteristic of the particles, and meanwhile construct the quantitative relations between the convection, diffusion coefficients and such mesoscopic probability parameters as blocking effect parameter of the percolation model as well as external potential parameter. In Section 3, Monte-Carlo numerical simulation will be given for verifying and modifying the mesoscopic particle migration theory. In Section 4, we will present the applicable qualification of the proposed theory, and obtain the relation between the blocking effect parameter and connecting probability of the percolation model.
Our conclusions will be given in Section 5.

%
%
%
%For providing a general theoretical foundation to the application of convection-diffusion equation,
%this study treat the motion of particle as a random walk and convection-diffusion equation under
%one dimensional situation is derived from this perspective. Relation between diffusion and
%convection coefficients with $p$ and $\alpha$, two parameters describe particle moving probability and
%blocking effect respectively, is gotten in this part. In third part, isotropic percolation configuration
%is used to represent porous media and the random walk is embedded into it. Theoretical result is
%confirmed by simulation, and results of convection and diffusion coefficients are extended to the
%entire interval of particle moving probability. As percolation probability $P$ represent connectivity
%of porous media and connectivity impact the transportation of particles, relation between $\alpha$ and $P$
%and blocking effect are discussed by simulation~\cite{}.

\vspace{5mm} \noindent {\Large{\bf 2. Mesoscopic particle migration theory}} \vspace{3mm}

%particle migration law

% The following hypotheses: theoretical
%(1) The interaction between/among particles is ignored;
%(2) The probability of the particles moving to which direction is determined by percolation configuration and external potential together;
%suppose that the interaction between/among particles is ignored. controllable
%The above hypotheses seem to be unreasonable, while our results show that we indeed focus the main points.  of particle migration in porous media
%Under this circumstance; As expected

In this part, based on the well-known bond percolation configuration, the particle migration in porous media is investigated from the mesoscopic perspective. In bond percolation, for a cubic lattice, every edge is occupied by a bond with a certain probability $P$ ($0\leq P\leq1$), and particles can only move on the bonds. In fact, if the particles move from a vertex of the cubic lattice to the connected bonds randomly, the isotropic bond percolation model comes into being. It is found that there is a percolation threshold $P_c$, and when $P>P_c$, the percolation phase transition appears so that the particles can percolate freely from one side of the isotropic bond percolation model to the opposite side. Apparently, when $P=1$, the isotropic bond percolation model comes to be fully connected.
%Motivated by the isotropic percolation model, researchers specify each bond a certain propagation direction and introduce the directed bond percolation model. The directed percolation model can present the directional propagation dynamics of the particles under certain conditions (such as epidemic dynamics). In both of the models, discrete
%, i.e., given a percolation configuration, the particles under external potentials cannot move along each bond with expected probability, hence, the particle migration behaviors cannot be well reflected.  discrete£¬ law
In the isotropic models, the probabilities of particles moving in a certain direction are completely determined by the bond distributions. In other words, if the particles are subjected to certain external potentials, such as thermal driving force, the particle migration behaviors would not be well reflected.

Hereby, we will consider the effect of unidirectional uniform external potentials on the particle behaviors, and base on the isotropic bond percolation configuration for exploring the particle migration under $P>P_c$. For simplicity, we begin from a symmetrical three-dimensional (3D) bond percolation configuration, which is not a strictly isotropic percolation model, as described in Fig.~\ref{gailvmodel}(a), and consider the particle behaviors under an upward uniform external potential. From the mesoscopic viewpoint, the particle migration is considered as a discrete motion, i.e., the total migration time is divided into several equal parts by a given time step $\tau^\ast$, and the particles migrate between the vertices of the percolation model and just right move a bond-length distance $\lambda$ within a time step $\tau^\ast$. Furthermore, the following hypotheses are presented:

%and just moves a bond-length distance $\lambda$ within a time step $\tau^\ast$. Besides, the following hypotheses are presented:
%  time
%the total particle migration time is divided into several equal parts by a given time step $\tau^\ast$, and the
%
%the particle migration is considered as a discrete motion, i.e., the particles are assumed to move between the vertices, and
%
%the particle migration time is divided into several equal parts by a given time step $\tau^\ast$, and
% given a time step $\tau^\ast$,
%, the total particle migration time is divided into several equal parts by a given time step $\tau^\ast$, and the particles are assumed to move through a bond length $\lambda$ just within $\tau^\ast$.
%Besides, the following hypotheses are presented:

%mesoscopic

\vspace{3mm}

(1) For a single particle, it possesses the diameter size of $10^{-9}\sim10^{-7}$m and keeps moving between the bond of the percolation model. Meanwhile, the interaction between/among the particles is ignored.

(2) When the percolation configuration becomes compact and complex, the particles move so fast from one end of a bond to the other that we only consider particle amounts at the vertices for computing the particle distribution in the percolation model.

(3) The effect of external potentials on the particle behaviors is all reflected in the probability of the particles moving from a vertex to each bond direction.

\vspace{5mm}
%(4) In a given time interval, the particles move from one end of a bond to the other end. passable
%, the particles move from one end of a bond to the other end.
%\scriptsize corresponding  at the next moment;  with different ratios;  performance; features; swarm
%the macroscopic 3D particle migration model for the particle swarm behavior is mapped into a mesoscopic 1D particle migration model for a single particle feature; a single particle; a series of
%here the horizontal movement of the particles is considered free and random; single

%the probability it keeps stationary is just the macroscopic ratio of all particles of the plane staying at this plane after time $\tau^*$,
%
%%the probability it keeps stationary is just the macroscopic ratio of the particles of the plane staying at the original plane after time t
%
%
%the probability it keeps stationary is just the macroscopic ratio of the corresponding particle swarm staying at the original plane after $\tau^*$,
%
%
%the probability it keeps stationary is just the macroscopic ratio of all particles of the plane staying in this plane after time $\tau^*$,
%The probability it keeps stationary is just the ratio of the number of in-plane moving particles to the total number of particles in the plane, associated
%the probability it keeps stationary is just the macroscopic ratio of all of the particles in the plane staying at the original plane after time $\tau$

\vspace{7mm}
\noindent\begin{minipage}{\textwidth}
\centering
\renewcommand{\captionfont}{\sffamily}
\renewcommand{\captionlabelfont}{\sffamily}
\renewcommand{\captionlabeldelim}{.\,}
\renewcommand{\figurename}{Fig.\,}
\includegraphics[scale=0.56]{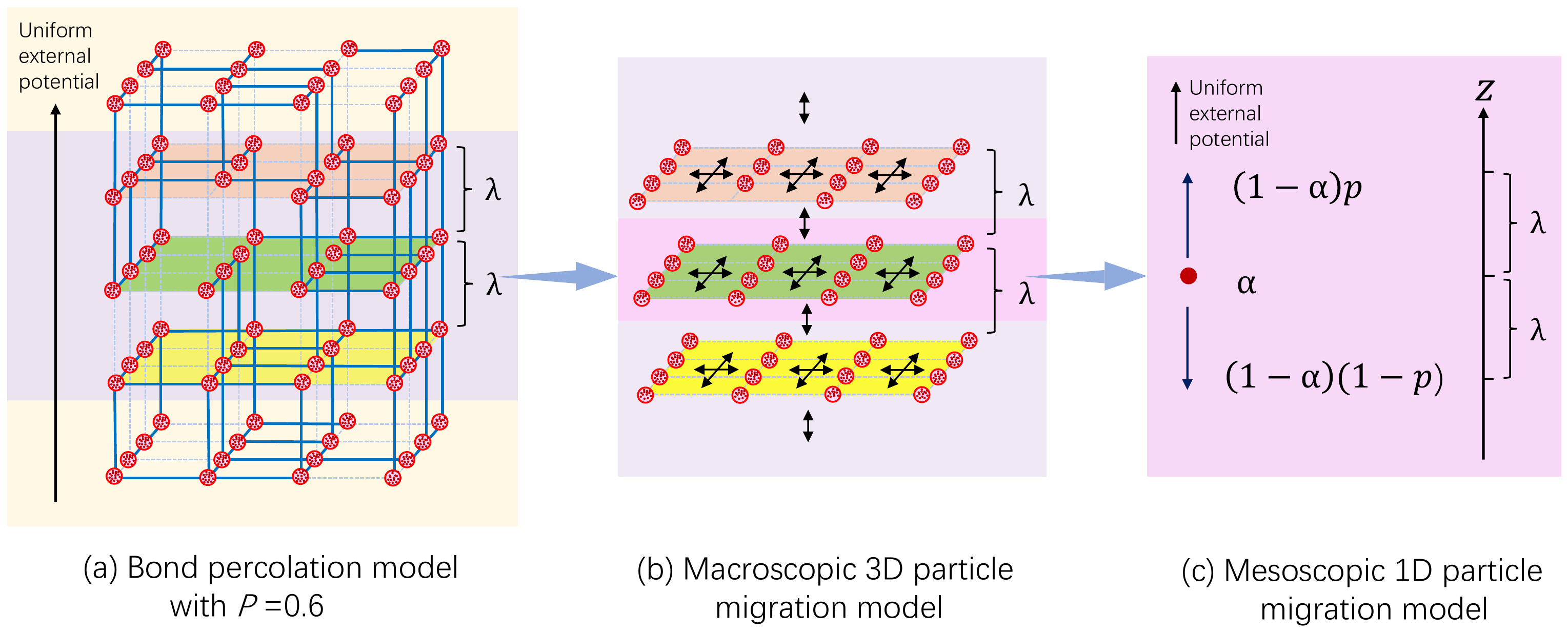}
{\center\figcaption{A 3$\times$3$\times$4 bond percolation configuration (\textit{P}=0.6) with particle migration and its discrete mesoscopic 1D particle migration model, in the figure $\lambda$ denotes the bond length, also the step length of the particles in a given time step $\tau^*$, \textsf{$\alpha$} denotes the probability of the particles staying in the original planes after $\tau^*$, and \textit{p} denotes the probability of the particles moving upward (also \textit{z} direction) under upward uniform external potential provided that they leave out of the original planes.
}
\label{gailvmodel}}
\end{minipage}
\vspace{5mm}

It is apparent that under the upward uniform external potential the particles will flow to the adjacent vertices through the available bonds after a time step $\tau^*$, and then all of the particles would be redistributed to all of the vertices of the percolation model.
%(here the movement of particles in the horizontal bond directions is considered free and random).
Taking all of the particles in a horizontal plane as a whole, we easily find that one part of the particles escapes from the original plane to the adjacent planes after $\tau^*$, while the other still stays at the original plane. From the macroscopic perspective, the particles are regarded as migrating inside and between the horizontal planes, as shown in Fig.~\ref{gailvmodel}(b). Furthermore, based on the quantitative ratios of the in-plane moving particle number and inter-plane moving particle number to the total number of particles of each plane, we may reconstruct the particle migration model from the mesoscopic perspective. As depicted in Figs.~\ref{gailvmodel}(b) and~\ref{gailvmodel}(c), taking one particle out of a plane and making a label, we can obtain its probabilistic migration characteristic at this time, specifically, the probability it keeps stationary is just the macroscopic ratio of all particles of this plane staying at the original plane after time $\tau^*$, which hereby is defined as blocking parameter and expressed as $\alpha$ ($0\leq \alpha\leq1$), and meanwhile its probability for moving upward (positive $z$ direction) and downward is $1-\alpha$.
Note that the label particle migrates upward and downward over time, and when it reaches another $z$ point, the blocking parameter $\alpha$ becomes different. In fact, although $\alpha$ varies with $z$ and is finally determined by the corresponding percolation model, while indeed
it can be regarded as invariant approximately when $P$ becomes large, i.e., the percolation configuration becomes compact and complex.
%indeed it keeps unchanged when $P$ is close to 1.
Under this circumstance, the mesoscopic 1D particle migration model is supposed to possess a constant $\alpha$ for all of $z$.
%Hereby we suppose that the mesoscopic 1D particle migration model possesses a constant $\alpha$ for all of $z$.
More important, with the consideration of the upward uniform external potential, the label particle moves upward and downward with different probabilities, and we express them respectively as $(1-\alpha)p$ and $(1-\alpha)(1-p)$, where $p$ denotes the effect of the upward external potential on the particle migration probability and $0\leq p \leq1$.

Taking $t$ as total particle migration time, we have $N=t/\tau^*$, and $N$ is the migration number of the label particle in time $t$. The direct result is that it keeps stationary $N_0=\alpha N$ times, moves upward $N_1=(1-\alpha)pN$ times and moves downward $N_2=(1-\alpha)(1-p)N$ times during $N$ migrations. Let $M$ denote the net number of the label particle moving in positive $z$ direction, and we have $M=N_1-N_2$. For a given integer $m$ ($m=0,1,2,\cdots, N$), the
probability of $M=m$ can be calculated as $\eta(m,N)=C^{N_1}_{N-N_0}p^{N_1}(1-p)^{N_2}$, where
$C^{N_1}_{N-N_0}$ denotes the number of combination. When $N$ is large, via the Stirling formula, $\eta(m,N)$ can be rewritten as
\begin{equation}\label{etamN}
\begin{aligned}
\ln \eta(m, N)=& \frac{1}{2} \ln \frac{2}{\pi}+(1-\alpha) N \ln (1-\alpha) N+(1-\alpha) N \ln 2+\frac{(1-\alpha) N}{2} \ln p(1-p)
\\&
+\frac{m}{2} \ln \frac{p}{1-p}-
\frac{(1-\alpha) N+m+1}{2} \ln [(1-\alpha) N+m]
\\&
-\frac{(1-\alpha) N-m+1}{2} \ln [(1-\alpha) N-m].
\end{aligned}
\end{equation}
Noticing that the probability of the particle at infinity is 0, we respectively expand $\ln[(1-\alpha)N+m]$ and $\ln[(1-\alpha)N-m]$ in Eq.~(\ref{etamN}) at their expectations $2(1-\alpha)Np$ and $2(1-\alpha)N(1-p)$ to quadratic terms and obtain
\begin{equation}\label{etamN2}
\begin{aligned}
\ln \eta(m, N)=&-\frac{1}{2} \ln 2 \pi-\frac{1}{2} \ln p(1-p)-\frac{1}{2} \ln (1-\alpha) N+\frac{(2 p-1)^2\left(6 p^2-6 p+1\right)}{16 p^2(1-p)^2}(1-\alpha) N+\\
& \frac{(2 p-1)^2\left(6 p^2-6 p+1\right)}{16 p^2(1-p)^2}-\frac{(2 p-1)(4 p-1)(4 p-3)}{16 p^2(1-p)^2} m+\frac{14 p^2-14 p+3}{16 p^2(1-p)^2(1-\alpha)} \frac{m^2}{N}.
\end{aligned}
\end{equation}
The corresponding position of the particle is $z=m\lambda$ when $M=m$. By substituting $m=\frac{z}{\lambda}$ and $N=\frac{t}{\tau^*}$ in Eq.~(\ref{etamN2}), after serializing and normalizing ($\int^{+\infty}_{-\infty}\eta dz =1$), we have continuous $\eta$ as
\begin{equation}\label{etamN3}
\eta(z, t)=\frac{1}{\sqrt{\pi}} \sqrt{\frac{k_1}{t}} \cdot \exp \left\{-\frac{k_1}{t}\left(z-\frac{k_2 t}{2 k_2}\right)^2\right\},
\end{equation}
where $k_1=-\frac{\left(14 p^2-14 p+3\right) \tau^*}{16 p^2(1-p)^2(1-\alpha) \lambda^2}, ~k_2=-\frac{(2 p-1)(4 p-1)(4 p-3)}{16 p^2(1-p)^2 \lambda}$.
In fact, the process of the particle moving to $z$ at time $t+\tau$ can be divided into two steps: first, the particle
moves to $z-\Delta$ at time $t$, and then takes place a displacement $\Delta$ during time $\tau$, which can be expressed as
\begin{equation}\label{etamNtaddtau}
\begin{aligned}
\eta(z, t+\tau) &=\int_{-\infty}^{+\infty} \eta(z-\Delta, t) \eta(\Delta, \tau) d \Delta=\int_{-\infty}^{+\infty}\left[\eta(z, t)-\Delta \frac{\partial \eta}{\partial z}+\frac{\Delta^2}{2} \frac{\partial^2 \eta}{\partial z^2}+\ldots\right] \eta(\Delta, \tau) d \Delta \\
&=\eta(z, t)-\frac{k_2 \tau}{2 k_1} \frac{\partial \eta}{\partial z}+\frac{\tau}{4 k_1} \frac{\partial^2 \eta}{\partial z^2}+o\left(\tau^\frac{3}{2}\right).
\end{aligned}
\end{equation}
Dividing both sides of Eq.~(\ref{etamNtaddtau}) by $\tau$ and letting $\tau\to 0$, we derive
\begin{equation}\label{kuosan}
\begin{aligned}
\frac{\partial \eta}{\partial t}= D \frac{\partial^2 \eta}{\partial z^2}- q \frac{\partial \eta}{\partial z},
\end{aligned}
\end{equation}
where $D=-\frac{4p^2(1-p)^2(1-\alpha)\lambda^2}{(14p^2-14p+3)\tau^*}$ and $q=\frac{(2p-1)(4p-1)(4p-3)(1-\alpha)\lambda}{2(14p^2-14p+3)\tau^*}$. It can be seen
hereby that Eq.~(\ref{kuosan}) is essentially a convection-diffusion equation, and the units of $D$ and $q$ are respectively m$^2/$s and m$/$s, agreeing with the dimension of the diffusion coefficient and convection coefficient obtained by predecessors.

Only one label particle's behavior is considered above. As Eq.~(\ref{kuosan}) is a linear equation, via the superposition principle, it is clear that Eq.~(\ref{kuosan}) can also be extended to describe the migration of all of particles in the macroscopic 3D percolation model. That is, if the total particle number in the percolation model is set as 1, $\eta(z,t)$ would be the vertical distribution function of particle concentration at any time. Surprisingly, in the present paper, instead of classical Fick's law, we reformulate the convection-diffusion equation from the mesoscopic bond percolation model, and obtain the relation between the macroscopic parameters $D,~q$ and mesoscopic parameters $\alpha,~p,~\tau^\ast,~\lambda$. In addition, note that $D,~q$ have two singular points $\frac{1}{14} \left(7\pm\sqrt{7}\right)$, and the reason is that only two terms in Eq.~(\ref{etamN2}) are retained in the Taylor expansion. In fact, the relations between $D,~q$ and $\alpha,~p,~\tau^\ast,~\lambda$ are only correct for local $p$ close to $\frac{1}{2}$, and in next part we will derive
global relations between $D,~q$ and $\alpha,~p,~\tau^\ast,~\lambda$ using the numerical simulation.

%That is, the particle migration in the bond percolation model is described by the convection-diffusion equation under certain condition.
%It is apparent that the particle migration in the bond percolation model is described by the convection-diffusion equation under certain condition. In addition,
%the reason why $D$ and $q$ have singular points is that only two terms in Eq.~(\ref{etamN2}) are retained in the Taylor expansion. In fact, in Eq.~(\ref{kuosan}), the relations between $D$ and $p$, $q$ and $p$ are only correct for local $p$, and global relations of $D-p$ and $q-p$ at $p\in[0,1]$ are derived by
%numerical simulation in the next part. the products of several factors,

\vspace{5mm} \noindent {\Large{\bf 3. Modification to the mesoscopic particle migration theory}} \vspace{3mm}

Under the upward potential, the theory of particle migration in symmetrical 3D bond percolation model has been presented when $P$ is large from the mesoscopic perspective, i.e., the convection-diffusion equation. As is known, convection-diffusion equation possesses analytical solutions, which help us test and verify the proposed theory through the comparison with numerical experiment result. Next, we will present the condition under which the theory applies, and meanwhile modify the expressions of diffusion coefficient and convection coefficient in the theory for expanding its applicability. Without loss of generality, hereby we will take $P=1$ into consideration for giving the result.

%As $D$ and $q$ both are not related to $P$,

When $P=1$, the bond percolation model becomes fully connected and $\alpha=\frac{2}{3}$. In the numerical experiment, we consider a $40\times 40\times 7000$ percolation configuration with the bond length $\lambda=1$, set the time step $\tau^\ast$ and total number of particles both as 1, and let them uniformly distribute to all vertices of the middlemost plane of the percolation model at the initial time $t=0$. Then with the consideration of the external potential $p$, the probabilities of the particles at non-boundary vertices moving upward, moving downward and staying at the original plane are respectively as $\frac{p}{3}$, $\frac{1-p}{3}$ and $\frac{2}{3}$, and the probabilities of the particles at boundary vertices should depend on the circumstances. Through the Monte-Carlo numerical simulation, when $t=3000$, we obtain the vertical particle distribution at $z=10$, $z=30$, $z=50$ under $p=\frac{1}{2}$ and $p=\frac{11}{20}$ respectively in Figs.~\ref{theorywithnumerical}(a) and (b). Theoretically, in this case, Eq.~(\ref{kuosan}) can be rewritten as
\begin{equation}\label{kuosanwithicbc}
\left\{
\begin{aligned}
%nomumber
&\frac{\partial \eta}{\partial t}= D \frac{\partial^2 \eta}{\partial z^2}- q \frac{\partial \eta}{\partial z},\\
&\eta|_{z\to \pm \infty}=0,~~~\eta(z,0)=\delta(0).\\
\end{aligned}
\right.
\end{equation}
 Eq.~(\ref{kuosanwithicbc}) has the following solution,
\begin{equation}\label{solutionkuosanwithicbc}
\begin{aligned}
\eta=\frac{1}{\sqrt{4\pi D t}}e^{-\frac{(z-qt)^2}{4Dt}},
\end{aligned}
\end{equation}
which is also depicted in Figs.~\ref{theorywithnumerical}(a) and (b). Fig.~\ref{theorywithnumerical} shows that theoretical results of the particle concentration at $z=10$, $z=30$ and $z=50$ all well match the numerical simulation results when $p=\frac{1}{2}$ and $p=\frac{11}{20}$, which verify the correctness of the theory. Unfortunately, when $p>\frac{11}{20}$, the theoretical results deviate rapidly from the numerical results, indicating that $D$-$p$ and $q$-$p$ relations only hold for a local range of $p$, to be specific, $\frac{9}{20}\leq p\leq\frac{11}{20}$. To modify the $D$-$p$ and $q$-$p$ relations so that they are applicable for global range of $p$, i.e., $0\leq p\leq 1$, motivated by the expressions of $D$-$p$ and $q$-$p$ in Eq.~(\ref{kuosan}), we rewrite them respectively as $D=S_D(p)\cdot B(\alpha)\cdot F_D(\lambda,\tau^*)$, $q=S_q(p)\cdot B(\alpha)\cdot F_q(\lambda,\tau^*)$, where $S_D$
and $S_q$ are factors only related to $p$ and reflect the statistical

\vspace{5mm}
\noindent\begin{minipage}{\textwidth}
\centering
\renewcommand{\captionfont}{\sffamily}
\renewcommand{\captionlabelfont}{\sffamily}
\renewcommand{\captionlabeldelim}{.\,}
\renewcommand{\figurename}{Fig.\,}
\includegraphics[scale=0.45]{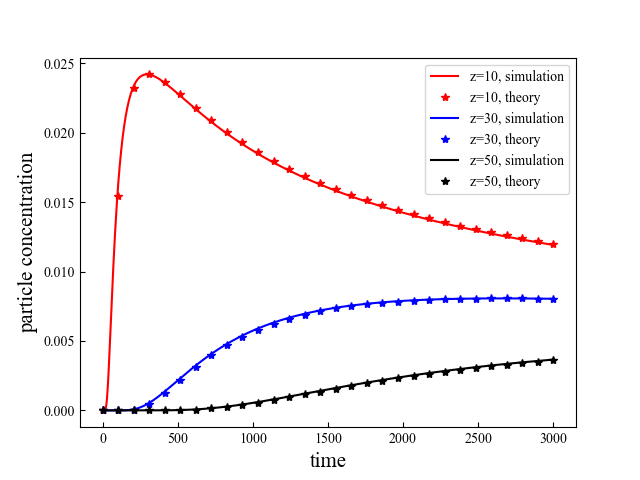}\hspace{0mm}\includegraphics[scale=0.45]{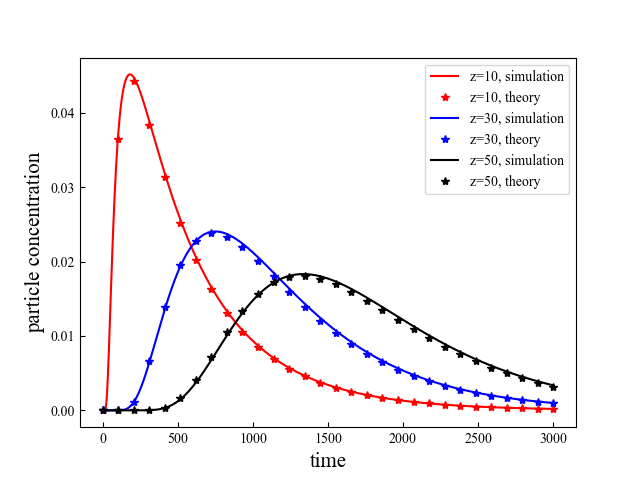}
{\center \textsf{(a)} \textsf{\textit{p}}=\textsf{1/2} \hspace{6cm} \textsf{(b)} \textsf{\textit{p}}=\textsf{11/20}}
{\center \figcaption{Comparisons between the theoretical and numerical results of the particle concentrations when \textit{p}=1/2 and \textit{p}=11/20.}
\label{theorywithnumerical}}
\end{minipage}

\vspace{5mm}
\noindent\begin{minipage}{\textwidth}
\centering
\renewcommand{\captionfont}{\sffamily}
\renewcommand{\captionlabelfont}{\sffamily}
\renewcommand{\captionlabeldelim}{.\,}
\renewcommand{\figurename}{Fig.\,}
\includegraphics[scale=0.5]{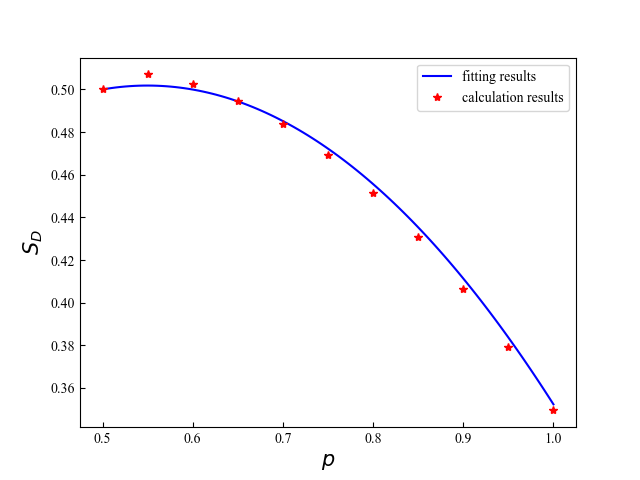}\hspace{0mm}\includegraphics[scale=0.5]{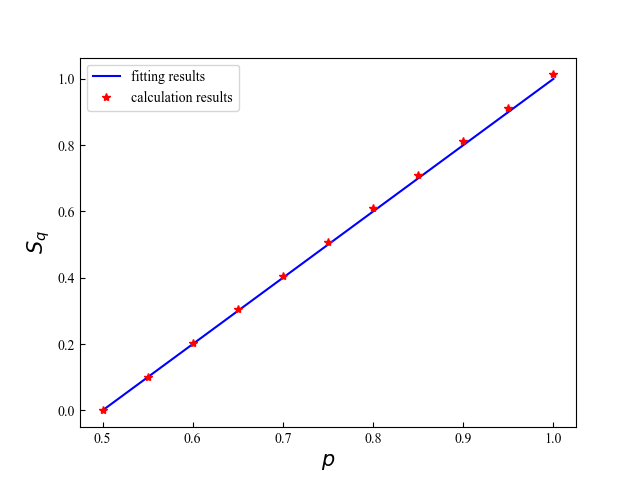}
\vspace{-3mm}
{\center \textsf{(a)} \textsf{\textit{S}}$_D$-\textsf{\textit{p}} \hspace{5.5cm} \textsf{(b)} \textsf{\textit{S}}$_q$-\textsf{\textit{p}}}
\vspace{-3mm}
{\center\figcaption{Fitting results of \textsf{\textit{S}}$_D$-\textsf{\textit{p}} and \textsf{\textit{S}}$_q$-\textsf{\textit{p}}.}
\label{SDandSq}}
\end{minipage}
\vspace{5mm}

\noindent characters of particles' mesoscopic motion, $B(\alpha)=1-\alpha$ reflects the blocking effects of porous media, $F_D=\frac{\lambda^2}{\tau^\ast}$ and $F_q=\frac{\lambda}{\tau^\ast}$ are related to $\lambda$, $\tau^*$ and reflect the kinetic characters of particles. To present the global expressions of $S_D$ and $S_q$, we use Eq.~(\ref{solutionkuosanwithicbc}) to fit the simulation results for obtaining $D$ and $q$ under different $p$.
Hereby we confine $p$ in $[\frac{1}{2},1]$ since $D$ and $q$ are respectively symmetrical and anti-symmetrical about $p=\frac{1}{2}$. Then, quadratic function is used to fit the relation between $S_D$ and $p$, and
linear function is used to fit relation between $S_q$ and $p$, as shown in Fig.~\ref{SDandSq}. With simple operations, we have
\begin{equation}\label{SD}
\begin{aligned}
%nomumber
&S_D= (p-1/2)(ap+b)+1/2,\\
%&S_D= (p-1/2)(-0.73713p+0.44179)+1/2,S_q= 2p-1
&S_q= 2p-1,\\
\end{aligned}
\end{equation}
where $a$ and $b$ are fitting coefficients, $a=-0.73713$, $b=0.44179$. Then, $D$ and $q$ in Eq.~(\ref{kuosan}) are modified as
\begin{equation}\label{realDQ}
\begin{aligned}
&D=\frac{\left[\left(p-\frac{1}{2}\right)\left(-0.73713p+0.44179\right)+\frac{1}{2}\right](1-\alpha)\lambda^2}{\tau^*},\\ &q=\frac{(2p-1)(1-\alpha)\lambda}{\tau^*}.
\end{aligned}
\end{equation}

%\begin{equation}\label{Sq}
%\begin{aligned}
%%nomumber
%\end{aligned}
%\end{equation}

\vspace{5mm} \noindent {\Large{\bf 4. Applicable qualification of the mesoscopic particle migration theory}} \vspace{3mm}

%Application condition
%Specific application condition for mesoscopic particle migration theory
%
%
%ÀíÂÛµÄÊÊÓÃÌõ¼þ
%Specific
%Moreover when $P$ is large, $\alpha$ can be approximately regarded as invariant for different $z$, and when $P=1$, the percolation model is fully connected so that $\alpha=\frac{2}{3}$.
%\vspace{5mm}\noindent {\large{\bf 3.3. Connectivity and blocking}}\vspace{3mm}

As known from above, $\alpha$ reflects the blocking effect during particle migration, and connecting probability $P$
reflects the connectivity of percolation configuration. Theoretically, when $P$ is larger than a certain value, $\alpha$ can be approximately regarded as invariant for different $z$ and under this circumstance the mesoscopic particle migration theory hold. Besides, as expected, blocking phenomenon becomes prominent when connectivity of the percolation configuration is low, hence there is a relation between $\alpha$ and $P$. Especially, when $P=1$, the percolation model is fully connected so that $\alpha=\frac{2}{3}$. In this part, we will explore the
relation between $\alpha$ and $P$, and meanwhile discuss the influence of $P$ on particle migration process via the numerical simulation. Note that $D=\frac{1-\alpha}{2}$ and $q=0$ when $p=\frac{1}{2}$ , in this case Eq.~(\ref{kuosan}) degenerates into the simple diffusion equation, and next discussion will be based on this situation.

In the Monte-Carlo numerical simulation, we use a $40\times 40\times 60$ percolation configuration, and set the bond length $\lambda$ and time step $\tau^\ast$ respectively as 1. Different from the initial condition used above, here we also set the total number of particles as 1, and let them distribute uniformly to all the vertices in the bottom plane of the percolation model at $t = 0$. As $p=\frac{1}{2}$ indicates that the probabilities of particles moving upward and downward are the same, in this case the particles move randomly in the percolation model. Next, with $L=60$, the theoretical equation converts to
\begin{equation}\label{kuosansim}
\left\{
\begin{aligned}
%nomumber
&\frac{\partial \eta}{\partial t}= D \frac{\partial^2 \eta}{\partial z^2},~~~z\in[0,L],\\
&\frac{\partial \eta}{\partial z}\bigg{|}_{z=0,L}=0, ~~\eta|_{t=0}=\delta(0),\\
\end{aligned}
\right.
\end{equation}
where $\delta(0)$ denotes the Dirac function. With simple mathematical operation, we have
\begin{equation}\label{solutionkuosansim}
\begin{aligned}
\eta(z,t)=\frac{1}{L} + \frac{2}{L} \sum\limits^{+\infty}_{k=1} \exp\left(-\frac{k^2\pi^2D}{L^2}t\right)
\cos\left(\frac{k\pi}{L}z\right).
\end{aligned}
\end{equation}
Then Eq.~(\ref{solutionkuosansim}) is used to fit the simulation results, and we can obtain $D$ at different $P$ values.
Fig.~\ref{simuandfitting} shows the fitting results when $P=0.9$ and $P=0.5$, and
it can be found that diffusion equation well describes the particle migration when $P>0.5$ by introducing the blocking parameter $\alpha$. Fig.~\ref{simuandcomparing} presents the simulation results of $P=0.4$ and $P=1$, and it is clear that when $P=0.4$ the particle concentration does not become homogeneous with time increasing, which indicates that the particles are blocked in certain areas of the percolation model. As it is a basic feature of Eq.~(\ref{solutionkuosansim}) that $\eta$ will tend towards homogeneous in space when $t$ tends to infinity, those simulation results indicate that essentially applicable qualification for the mesoscopic particle migration theory is $P\geq0.5$.

As the blocking parameter $\alpha$ can be calculated using $\alpha=1-2D$. Hence, when $P\geq0.5$, based the theoretical result that $\alpha=2/3$ when $P=1$, the quadratic function is chosen to fit the calculation results for obtaining the relation between $\alpha$ and $P$ as
\begin{equation}\label{relationofPalpha}
\begin{aligned}
\alpha=(P-1)(cP+d)+2/3
\end{aligned}
\end{equation}
where $c$ and $d$ are fitting coefficients, and $c=0.4821$ and $d=-0.5536$. The relation between $\alpha$ and $P$ is showed in Fig.~\ref{palpha}.

\noindent\begin{minipage}{\textwidth}
\centering
\renewcommand{\captionfont}{\sffamily}
\renewcommand{\captionlabelfont}{\sffamily}
\renewcommand{\captionlabeldelim}{.\,}
\renewcommand{\figurename}{Fig.\,}
\includegraphics[scale=0.5]{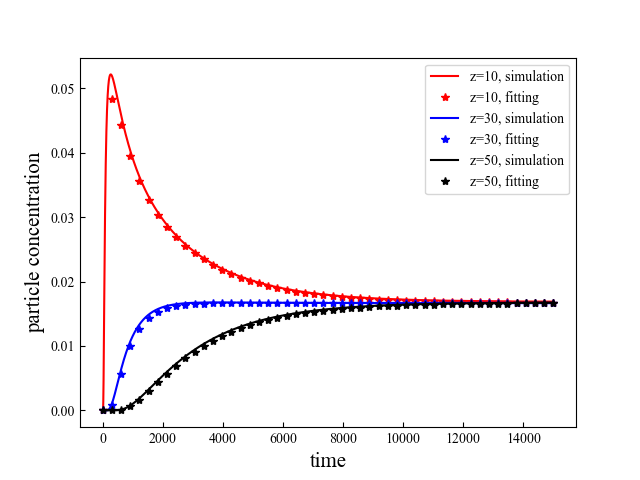}\hspace{0mm}\includegraphics[scale=0.5]{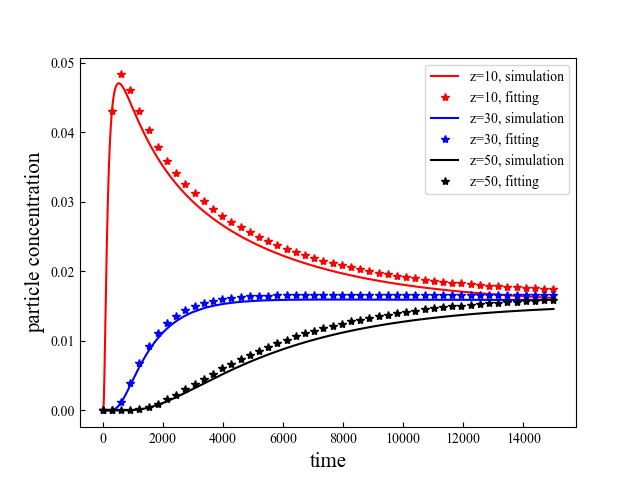}
{\center \textsf{(a)} \textsf{\textit{P}}=\textsf{9/10} \hspace{6cm} \textsf{(b)} \textsf{\textit{P}}=\textsf{1/2}   }
{\center
\figcaption{Simulation and fitting results of particles migration when \textsf{\textit{P}}$\neq$1.}
\label{simuandfitting}}
\end{minipage}

\noindent\begin{minipage}{\textwidth}
\centering
\renewcommand{\captionfont}{\sffamily}
\renewcommand{\captionlabelfont}{\sffamily}
\renewcommand{\captionlabeldelim}{.\,}
\renewcommand{\figurename}{Fig.\,}
\includegraphics[scale=0.5]{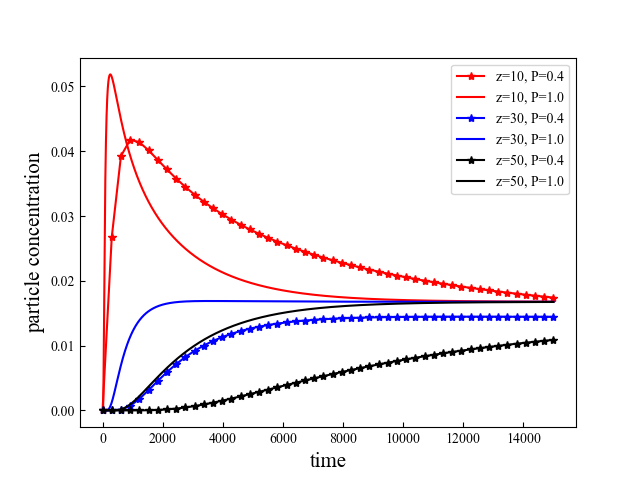}
{\center
\figcaption{Comparison of particles migration when \textsf{\textit{P}}=1 and \textsf{\textit{P}}=2/5.}
\label{simuandcomparing}}
\end{minipage}
\vspace{5mm}

\noindent\begin{minipage}{\textwidth}
\centering
\renewcommand{\captionfont}{\sffamily}
\renewcommand{\captionlabelfont}{\sffamily}
\renewcommand{\captionlabeldelim}{.\,}
\renewcommand{\figurename}{Fig.\,}
\includegraphics[scale=0.5]{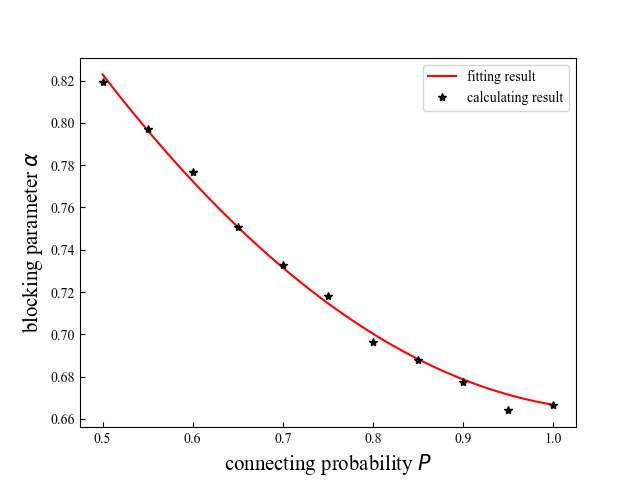}
{\center
\figcaption{Relation between \textsl{P} and $\alpha$.}
\label{palpha}}
\end{minipage}

\vspace{7mm} \noindent{\Large{\bf 5. Conclusions}}\vspace{3mm}

Particle migration in porous media has been studied from the mesoscopic level in this research and three
conclusions are drawn:

(1) Particle motion in mesoscopic level can be treated as a random walk and that
corresponding to diffusion-convection equation in macroscopic level. Diffusion and convection
coefficients can be written as $D=S_D(p)\cdot B(\alpha)\cdot F_D(\lambda,\tau^*)$ and $q=S_q(p)\cdot B(\alpha)\cdot F_q(\lambda,\tau^*)$. Relations between $S_D$, $S_q$ and $p$ have been gotten by theory and simulation as $S_D= (p-1/2)(-0.73713p+0.44179)+1/2$ and $S_q= 2p-1$;

(2) Percolation theory is implemented to study the influence of connectivity on the particle
migration. Diffusion-convection equation can be used to describe particle
migration in porous media only if the connecting probability of percolation model $P\geq 0.5$. Under this circumstance, the relation between blocking parameter $\alpha$ and $P$ is
gotten by simulation as $\alpha=(P-1)(0.4821P-0.5536)+2/3$;

% by introducing blocking parameter $\alpha$ to reflect the influence of blocking effect;
%obvious

(3) Blocking effect becomes manifest and its influence on particle migration tends to be non-linear
when $P<0.5$. Diffusion-convection equation is not applicable in this situation.

\vspace{10mm} \noindent {\Large{\bf Author declarations}}\vspace{3mm}

\textbf{Conflict of interest}: The authors have no conflicts to disclose.

\textbf{Data availability}: The data that support the findings of this study are available within the
article.

\vspace{10mm} \noindent {\Large{\bf Acknowledgements}}\vspace{3mm}

%We express our sincere thanks to the referees for their valuable comments. We are also grateful for technical support from the High Performance Computing Center of Central South University.
%This work has been supported by the National Key R\&D Program of China under Grant No. 2017YFE0119500, the Huxiang
%high-level talent gathering project (innovation team project) under Grant No. 2019RS1008, the Natural Science Foundation of Hunan Province of China under Grant No. 2022JJ40566, the Fundamental Research Funds for the Central Universities of Central South University under Grant 2022zzts0018, and the Outstanding Research Project of Shen Yuan Honors College, BUAA, under Grant No. 230121104.

We express our sincere thanks to the referees for their valuable comments.
%We are also grateful for technical support from the High Performance Computing Center of Central South University.
This work has been supported by the National Natural Science Foundation of China under Grant No. 52208378,
the Natural Science Foundation of Hunan Province of China under Grant No. 2022JJ40566.
%, the Fundamental Research Funds for the Central Universities of Central South University under Grant 2022zzts0018, the Hunan Provincial Postgraduate Research and Innovation Project under Grant No. CX20220109, and the Outstanding Research Project of Shen Yuan Honors College, BUAA, under Grant No. 230121104.

\end{document}